\documentclass{article}

\usepackage{arxiv}

\usepackage[utf8]{inputenc} 
\usepackage[T1]{fontenc}    
\usepackage{hyperref}       
\usepackage{url}            
\usepackage{booktabs}       
\usepackage{amsfonts}       
\usepackage{nicefrac}       
\usepackage{microtype}      
\usepackage{lipsum}

\usepackage{booktabs}
\usepackage{comment}
\usepackage[table,xcdraw,dvipsnames]{xcolor}
\usepackage{float}
\usepackage{textcomp}
\usepackage{color}
\usepackage{lettrine}

\usepackage{setspace}
\usepackage{amsmath}
\usepackage{amssymb}
\usepackage{graphicx}
\usepackage{moreverb}
\usepackage{siunitx}
\usepackage {tikz}
\usepackage{cleveref}
\usepackage{multirow}
\usepackage{algorithm}
\usepackage{algcompatible}

\algnewcommand\algorithmicto{\textbf{to}}
\algnewcommand\RETURN{\State \textbf{return} }

\usepackage{array}
\usepackage{mathtools}
\DeclarePairedDelimiter{\abs}{\lvert}{\rvert}
\usepackage[noadjust]{cite}

\title{DVNet: A Memory-Efficient Three-Dimensional CNN for Large-Scale Neurovascular Reconstruction}

\author{
  Leila~Saadatifard\\
  Department of Electrical and Computer Engineering\\
  University of Houston\\
  Houston, TX 77004 \\
  \texttt{lsaadatifard@uh.edu} \\
   \And
 Aryan~Mobiny \\
  Department of Electrical and Computer Engineering\\
  University of Houston\\
  Houston, TX 77004 \\
  \And
  Pavel~Govyadinov \\
  Department of Electrical and Computer Engineering\\
  University of Houston\\
  Houston, TX 77004 \\
  \And
  Hien V.~Nguyen \\
  Department of Electrical and Computer Engineering\\
  University of Houston\\
  Houston, TX 77004 \\
  \And
  David~Mayerich \\
  Department of Electrical and Computer Engineering\\
  University of Houston\\
  Houston, TX 77004 \\
  \texttt{mayerich@uh.edu}
}

\begin{document}
\maketitle
\begin{abstract}
Maps of brain microarchitecture are important for understanding neurological function and behavior, including alterations caused by chronic conditions such as neurodegenerative disease. Techniques such as knife-edge scanning microscopy (KESM) provide the potential for whole organ imaging at sub-cellular resolution. However, multi-terabyte data sizes make manual annotation impractical and automatic segmentation challenging. Densely packed cells combined with interconnected microvascular networks are a challenge for current segmentation algorithms. The massive size of high-throughput microscopy data necessitates fast and largely unsupervised algorithms. In this paper, we investigate a fully-convolutional, deep, and densely-connected encoder-decoder for pixel-wise semantic segmentation. The excessive memory complexity often encountered with deep and dense networks is mitigated using skip connections, resulting in fewer parameters and enabling a significant performance increase over prior architectures. The proposed network provides superior performance for semantic segmentation problems applied to open-source benchmarks. We finally demonstrate our network for cellular and microvascular segmentation, enabling quantitative metrics for organ-scale neurovascular analysis.
\end{abstract}


\section{Introduction}
The relationships between brain cells and microvessels are critical to brain function \cite{andreone2015neuronal, tsai2009correlations} and a frequent target in neurodegenerative research \cite{iadecola2004neurovascular}. However, disease-induced microstructural changes occur across large tissue regions, making comprehensive studies difficult due to limitations in microscope fields of view and imaging speed.

High-throughput microscopy aims to produce three-dimensional images of organ-scale tissue samples ($>$ \SI{1}{\centi\meter^3}) at sub-cellular resolution ($<$ \SI{1}{\micro\meter^3}). Two such techniques, knife-edge scanning microscopy (KESM) \cite{mayerich2008knife} and micro-optical sectioning tomography (MOST) \cite{li2010micro}, produce images containing hundreds of millions of cells alongside interconnected microvascular networks. While the resulting images have sub-micrometer spatial resolution, they are densely packed with a variety of microstructures, making manual annotation impractical and confounding automated algorithms. There is therefore a compelling need for tools to automate segmentation on massive images, making high-throughput microscopy practical for quantitative biomedical studies.

\begin{figure*}[t]
\centering
\includegraphics[width=\linewidth]{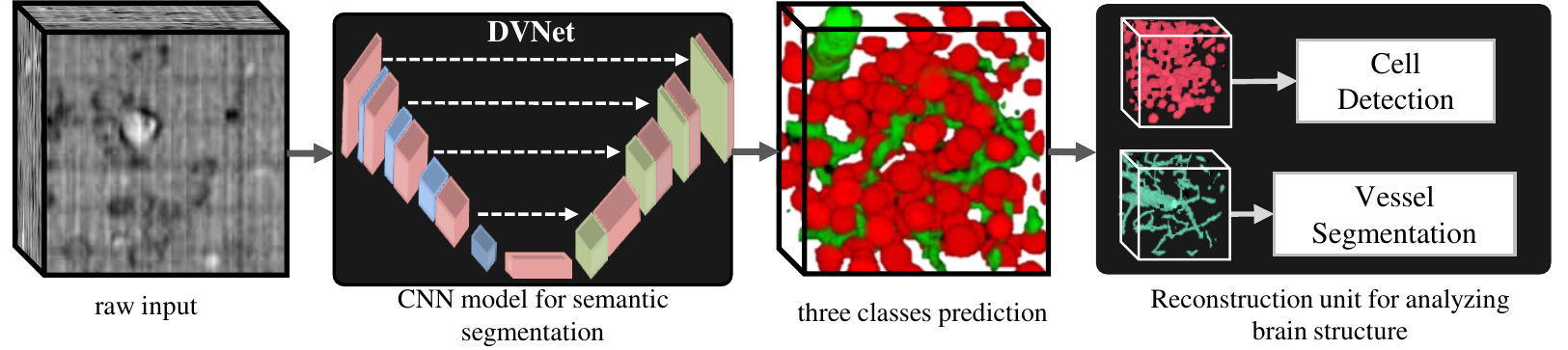}
\caption {The input volume is a stack of low contrast images of cellular, vascular, and neuropil structure from the mouse brain. DVNet is a fully convolutional encoder-decoder network using dense connections for pixelwise classification. It predicts different segmentation masks for all classes in the input to improve the contrast of the raw dataset. A GPU based postprocessing unit is applied on prediction masks from DVNet to reconstruct cellular and vascular structure. This unit is composed of two blocks for cell detection and vessel segmentation that are applied on cellular and vascular masks. This scalable framework  allows to analyze the whole mouse brain architecture.}  
\label{fig_model_config}
\end{figure*}

In this paper we introduce a framework for automated nuclear and microvascular segmentation in KESM images based on an efficient and accurate deep encoder-decoder architecture (Fig. \ref{fig_model_config}). KESM and MOST images are of particular interest in neurovascular anatomy \cite{mayerich2011fast,xiong2017precise} and have far-reaching applications for quantifying diseased tissue, including cancer grading and treatment efficacy. We propose two contributions critical to cellular and microvascular reconstruction at the macroscopic (\SI{1}{\centi\meter^3}) scale:
\begin{enumerate}
  \item A memory efficient densely-connected convolutional neural network for 3D semantic segmentation. A fully-convolutional encoder-decoder network is modified with dense connections and transition factors to control the number of trainable parameters and mitigate ``memory explosion.''  
  \item The cellular and vascular structure for a \SI{1}{\milli\meter^3} region is automatically segmented, demonstrating reliability, scalability, and accuracy exceeding prior methods.
\end{enumerate}

We first present related work in Section \ref{related_work}. Section~\ref{methodology} describes the proposed network architecture, training regime, and post-processing methods. We then discuss details for leveraged data sets in Section~\ref{sec_KESM_dataset} and experimental results are presented in Section~\ref{results}. Section~\ref{conclusion} concludes with a summary. Constraints and future improvements are discussed in Section~\ref{discussion}.

\section{Related Work}
\label{related_work}

Nuclear segmentation is challenging due to the wide variability in size, morphology, and density. It is often a multi-step process, relying on seed selection such as the \textit{Laplacian of Gaussian} \cite{kong2013generalized}, followed by segmentation steps, including watershed \cite{zhang2010neutrosophic} and graph-cut \cite{al2010improved} algorithms.

Automated seed selection for KESM and MOST data has been proposed using three-dimensional iterative voting \cite{saadatifard2016three, saadatifard2018robust}, however the complexity of chromatin structures limits accuracy. Staining heterogeneity and resolution limitations further reduce accuracy, causing over-seeding in poorly stained areas and under-seeding in dense cortical regions. 

Vessel segmentation is challenging given the complexity of the embedded capillary network. Prior work uses a combined template-matching and predictor/corrector approach for KESM and MOST data \cite{govyadinov2018robust}, however these algorithms are less accurate with nuclear stains \cite{xiong2017precise} (Fig. \ref{fig_kesm_nissl}), which have lower contrast than vascular perfusion labels and exhibit additional structural complexity from chromatin features in the nucleus.

Deep neural networks (DNNs) have gained attention by outperforming state-of-the-art algorithms in pattern recognition \cite{schmidhuber2015deep}, and may be amenable to high-throughput microscopy. These models use a cascade of processing stages to learn abstract representations from high-dimensional data \cite{lecun2015deep}. Convolutional neural networks (CNNs) were initially introduced for image processing \cite{lecun1995convolutional} and demonstrated excellent performance in signal processing \cite{hershey2017cnn, ravindran2019assaying}, including computer vision in images \cite{krizhevsky2012imagenet} and video \cite{karpathy2014large}. CNNs have been successfully applied to a wide range of tasks including classification \cite{wang2016cnn, berisha2019deep, egmont2002image, shahraki2018graph}, object detection \cite{girshick2015fast, foroozandeh2017cyclist}, digital staining \cite{lotfollahi2019digital}, and semantic segmentation \cite{milletari2016v}, and have been adopted for medical applications including x-ray pathology \cite{rajpurkar2017chexnet} and computed tomography \cite{liao2019evaluate, mobiny2017lung}. 

While powerful, CNNs have fundamental limitations \cite{mobiny2019automated}. First, they require a substantial number of training samples to generalize, making data collection and annotation an expensive and time-consuming barrier. Many modern architectures rely on millions of parameters, leading to exponential increases in computational and memory requirements. These architectures are therefore impractical for training with 3D images. \textbf{There is therefore a compelling need for new architectures that provide top-tier performance while minimizing trainable parameters.}

\subsection{CNNs for Semantic Segmentation}
Semantic segmentation is a critical component of biomedical image analysis. Adding additional convolutional layers to produce deeper networks has achieved progressively better performance \cite{simonyan2014very}, however the vanishing gradient problem is a well-known issue for deep architectures \cite{glorot2010understanding, srivastava2015training}. Residual networks were introduced \cite{he2016deep} to mitigate the vanishing gradient problem and reduce convergence time. Alternatively, dense blocks \cite{huang2017densely} alleviate the vanishing gradient problem by using feed-forward connections between convolution layers. Dense connections also reuse features, reducing trainable parameters.

Fully-convolutional encoder-decoder neural networks, which map input data to an identically-sized output \cite{goodfellow2016deep}, are a pioneering method for semantic segmentation \cite{ronneberger2015u}. These networks often using skip connections to copy features from the encoder to decoder path. This architecture has been extended for three dimensional data by adding residual connections at each stage \cite{milletari2016v, cciccek20163d}. The Tiramisu network \cite{jegou2017one} is a fully-convolutional dense network providing state-of-the-art performance for urban scene benchmarks \cite{BrostowFC:PRL2008} by using dense blocks in the encoding path to extract semantic features. However, dense blocks in \textit{both} encoder and decoder paths result in memory explosion. To address this problem, Tiramisu removes short connections and does not use a fully connected dense block. However, short connections convey useful information and do not add additional parameters \cite{he2016deep}. We propose a fully convolutional auto-encoder using fully connected dense blocks in both the encoder and decoder paths. Two transition factors are used to reduce feature scaling for large three-dimensional images.

\subsection{Cellular and Microvascular Segmentation Models}

Three-dimensional microvascular modeling is challenging \cite{bullmore2009complex} because most microscopes have a very limited field of view (\textless\SI{1}{\milli\meter}).
The distribution of cells relative to the microvascular network is key to learning about brain haemodynamics and neurovascular architecture for both diseased and healthy tissue. Quantifying their correlation is therefore crucial to understanding brain function and metabolism \cite{kleinfeld2011guide}. The regional density of cells and their distribution relative to the vascular network is important for understanding the efficiency and spatial localization of neurovascular reactions \cite{tsai2009correlations}. Recent advances in imaging provide high resolution datasets capturing detailed information about neuronal and non-neuronal cell distributions and microvessels from the whole rodent brain. However, researchers lack a reliable set of algorithms to segment cells and vessels from complex and low-contrast thionine-stained data \cite{tsai2009correlations, wu20143d}. The complexity of brain architecture at sub-micrometer resolution results in acquiring a terabyte-scale volumetric data. Data at this scale is particularly challenging to both manage and process due to the memory limitations and the computational complexity of existing imaging algorithms \cite{akil2011challenges}. To address these issues, a fast and efficient framework is required that precisely maps the whole brain from raw Feulgen-stained three-dimensional samples. Although it is known that cell size and densities vary across cortical and subcortical regions, there are no specific values that accurately quantify these values across the brain \cite{keller2018cell}.\\
Recent work uses expansion microscopy and tissue clearing to image the cellular structure of the mouse brain, including quantitative cell counting across brain regions \cite{murakami2018three}. Although some studies that measure cell densities in different mouse brain regions, there is substantial variation in the reported values. Further, there is more uncertainty about the densities of specific cell types in brain regions \cite{keller2018cell}. Most of the knowledge of regional cell density is estimated using the stereology method  \cite{mouton2013neurostereology}, therefore a reliable approach is required to quantify cell densities in the mouse brain.\\
Recent work on microvascular modeling has focused on light-sheet and multi-photon microscopy \cite{lauwers2008morphometry, blinder2013cortical, haft2019deep}. These methods are able to achieve voxel sizes of \SIrange[range-phrase = --]{1}{3}{\micro\meter} allowing reconstruction of arterioles and venules, as well as the capillary network, but are not able to precisely model the holistic 3D structure of the whole brain due to limited resolution or time constraints. Although capillaries play a vital role in neurovascular function, a comprehensive model of their network properties is unavailable \cite{smith2019brain}. Given the availability of data sets comprising large regions of the brain \cite{mayerich2011fast,xiong2017precise}, this limitation is primarily due to the difficulty segmenting and analyzing quantitative DNA-labeled brain images.\\
KESM images provide sufficient detail to quantify cellular and vascular structure and distributions. The tissue was stained using thionine perfusion \cite{choe2011specimen}, which acts as a nissl or Feulgen stain primarily labeling nucleic acids. The dye is sufficiently quantitative to differentiate between cell nuclei and chromatin with a dark purple stain. The surrounding neuropil is lightly labeled, while vessels and microvessels contain no tissue and are therefore unstained. The neuropil labeling allows differentiation between microvessels and cell nuclei. However, the low contrast difference between the labels, as well as the complex structure of the cell nucleus, makes cell localization and vascular segmentation with standard techniques challenging. Semantic segmentation forms the basis for explicit neurovascular modeling by classifying pixels into cellular, vascular, and background components.

\section{Methodology}
\label{methodology}
We develop a deep fully-convolutional network that efficiently combines long skip connections proposed in U-Net \cite{ronneberger2015u} with dense connections (short connections) proposed in the DenseNet CNN family \cite{huang2017densely}. Long skip connections are typically used to pass feature maps from encoder to decoder, recovering spatial information lost during downsampling \cite{cciccek20163d}. Short connections directly connect feature maps from preceding layers within a block, creating shortcuts for uninterrupted gradient flow \cite{huang2017densely}. However, feature map concatenation through dense connections can quickly introduce memory explosion. We therefore introduce a trainable feature compression mechanism to constrain information flow. 

We tested both two-dimensional and three-dimensional architectures, with two-dimensional performance tested on standard data sets. Our three-dimensional extension is used to differentiate cellular and microvascular structures from background in KESM images. Finally, the outputs of the CNN (i.e. segmentation masks) undergo post-processing algorithms to create the cellular and vascular model. We show that the cascade of semantic segmentation step and the employed post-processing algorithms provides a significant improvement in model accuracy over existing algorithms used to process KESM images.

\subsection{Dense-VNet (DVNet)}
\begin{figure*}[t]\centering
\includegraphics[width=\linewidth]{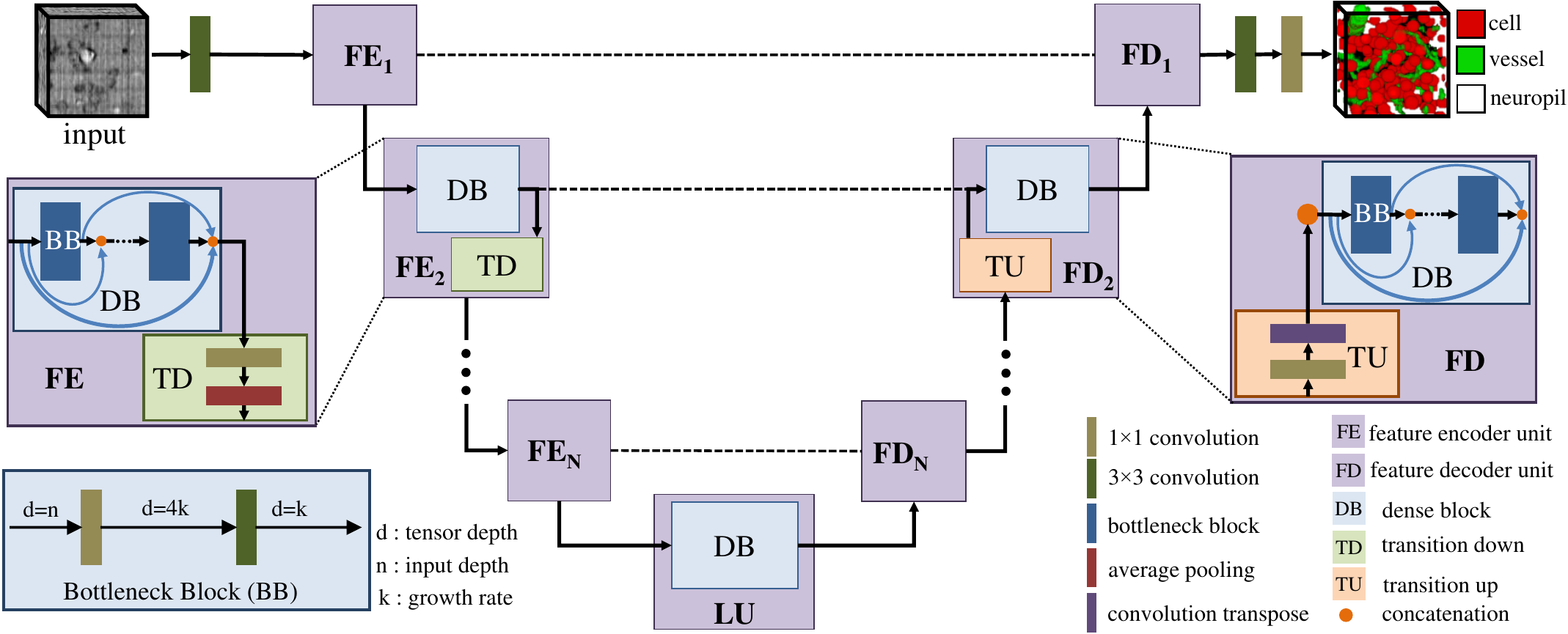}
\caption{The proposed network architecture (DVNet) consists of an encoder, a decoder and a linking unit. Feature extraction units form encoder and decoder (FE and FD). These units include a dense block and a transition block. Dense blocks are residual networks that compute deep features making from bottleneck blocks (BB). The bottleneck block as shown in this figure is made of two convolutional layers to extract the information with the fixed depth (k). The transition down block (TD) in the encoder decreases the spatial resolution for the next level. The transition up block in the decoder upsamples spatial dimension to reconstruct the output.}
\label{fig_net_arch}
\end{figure*}

The DVNet architecture contains densely-connected encoder and decoder paths joined by a \textit{linking unit} (Fig. \ref{fig_net_arch}). The encoder path decomposes the input image into a feature hierarchy, while the decoder reconstructs spatial context while performing per-pixel classification. The linking unit (LU) provides the primary information path and contains the most compact representation of the input. Cascades of feature encoder (FE) and feature decoder (FD) units run in parallel to form the encoder and decoder paths respectively (see Fig. \ref{fig_net_arch}).

\vspace{2mm}
\noindent
\textbf{Dense Block (DB):} is a sequence of multiple bottleneck blocks with direct short connections from any layer to all subsequent layers (referred to as dense connections). Feature maps for layer $\ell$ are computed by:
\begin{equation}
f_l = H_l\left(\left[f_0, f_1, \dotsb, f_{l-1}\right]\right).
\label{eq_dense_connection}
\end{equation}
where $f_i$ is the feature map for layer $i$ in the dense block, $[f_0, f_1, \dotsb, f_{l-1}]$ refers to the concatenation of feature maps of layers 0 to $l-1$, and $H_l(.)$ calculates the feature map for layer $\ell$. Dense connection of the feature maps strengthens feature propagation, mitigates the vanishing gradient problem, and applies a regularizing effect to reduce over-fitting \cite{huang2017densely, drozdzal2016importance}. In a dense block, short skip connections pass input features to the output (bold blue lines in Fig. \ref{fig_net_arch}).

\vspace{2mm}
\noindent
\textbf{Bottleneck block (BB):} is the main component of a dense block. Each BB is composed of two convolutional layers: a $1\times1$ convolution layer with $4k$ convolution filters followed by a $3\times3$ layer with $k$ filters, where $k$ is a hyper-parameter named as \textit{growth rate}. The $1\times1$ convolution serves as the bottleneck layer which reduces the number of input feature maps, and thus increases the computational efficiency \cite{huang2017densely}. Note that a batch normalization and ReLU nonlinearity (Fig. \ref{fig_net_arch}) precedes each convolutional layer.

A DB with $L$ layers and a growth rate of $k$ calculates $\left(f_0 + L\times k\right)$ feature maps, where $f_0$ is the input depth. Feature maps computed at each encoder level are concatenated to the inputs of the DB in the corresponding level in the decoder path using a long skip connection (the dashed lines in Figure \ref{fig_net_arch}). Long skip connections restore the spatial context lost through encoder down sampling \cite{long2015fully, drozdzal2016importance}, while short skip connections alleviate vanishing gradients and improve convergence \cite{he2016deep, szegedy2017inception}. 

\vspace{2mm}
\noindent
\textbf{Feature Compression: }
Employing both the short and long skip connections together rapidly increases the number of concatenated feature maps and causes memory explosion. In Tiramisu network \cite{jegou2017one}, the short skip connections of the decoder path were removed so as to decrease the number of feature maps and mitigate the explosion. This, however, leads to information loss, and may cause vanishing gradients which eventually degrades the convergence and prediction accuracy performance. In our proposed architecture, DVNet, we introduce compression factors which helps controlling the number of feature maps as explained below.

\vspace{2mm}
\noindent
\textbf{Transition block:} is used along with a dense block to form a feature extraction unit (see Fig. \ref{fig_net_arch}) and aims to change the size of the feature maps. Transition down (TD) blocks use two convolution layers to downsample the feature maps by $2\times$ at each level of the encoder path. Conversely, Transition Up (TU) blocks are used at each level of the decoder path to upsample the feature maps by $2\times$ using two transposed-convolution layers \cite{zeiler2010deconvolutional}. To further improve the model compactness, we introduce two transition factors, namely $\theta_D$ and $\theta_U$, to control the number of feature maps in the encoder and decoder paths respectively. If the output of a DB contains $M$ feature maps, the following TD (TU) block will generate $\lfloor\theta_DM\rfloor$ ($\lfloor\theta_UM\rfloor$) output feature maps using a $1\times1$ convolution layer, where $0<\theta_D, \theta_U \leq1$ and $\lfloor.\rfloor$ represents the floor function.

The \textbf{encoder} path starts with a $3\times3$ convolution layer, followed by $N$ feature encoding (FE) blocks. Coarse information is extracted at the deepest encoder layers with broader receptive fields, while shallow blocks compute local information. Hierarchical representations are stored in dense block outputs and compressed through average pooling in TD blocks. Note that BN and ReLU operate on inputs for all convolutional layers in the encoder path.

The \textbf{linking unit} that joins encoder and decoder paths is composed of a single dense block and generates deep features with low spatial specificity.

The \textbf{decoder} receives semantic features at each spatial resolution and reconstructs pixel-wise masks for each class with the same spatial context as the input image. The decoder is composed of $N$ feature extraction blocks, followed by two convolutional layers. In each TU block, a $1\times1$ convolution decreases the feature map depth by $\theta_U$ factor. Then a $3\times3$ transposed convolution layer upsamples the spatial size by 2. Finally, the output of the TU block is concatenated with the output from the DB of the corresponding encoder to form the input of the succeeding DB. At the end of the decoder path, two $3\times3$ and $1\times1$ convolution layers operate on the feature maps computed by the decoder to generate the output masks for the semantic classes.


\subsection{Hyperparameters}

The final DVNet consists of 5 feature extraction levels (FEs and FDs) composed of increasing numbers of bottleneck blocks $\left(4, 6, 8, 10, 12\right)$. The linking unit has $16$ layers, and the growth rate for the network is 16. Transition down and transition up factors are $0.5$ and $0.3$. The output mask has the same size as the input with the number of channels equals to the number of predicted classes. The input layer computes a fixed number feature maps (64) while having the same size as input ($X$). Feature map depth are expanded and its dimension are halved when passing each level of the encoder. In the decoder path, its dimension is upsampled to finally generate $X$ size masks. The feature map depth is also accumulated but the $\theta_{\uparrow}$ prevents it from explosion (Table \ref{table_arch_spec}). 

\begin{table}[t]
\caption{Network architecture and feature map sizes for the best performing network, DVNet-v3. It contains 5 feature extraction units (FEs and FDs) with $\theta_D=0.5$ and $\theta_U=0.3$. The growth rate is set to $k=16$.}\centering
\begin{tabular}{|l|c|c|}
\hline
layer name      & \multicolumn{1}{l|}{feature depth} & \multicolumn{1}{l|}{feature dimension} \\ \hline
input conv      & 64                               & X                                      \\ \hline
DB(4 BBs)       & 128                              & X                                      \\ \hline
TD + DB(6 BBs)       & 160                              & X/2                                    \\ \hline
TD + DB(8 BBs)       & 208                              & X/4                                    \\ \hline
TD + DB(10 BBs)      & 264                              & X/8                                    \\ \hline
TD + DB(12 BBs)      & 324                              & X/16                                   \\ \hline
TD + LU(16 BBs)      & 418                              & X/32                                   \\ \hline
TU + DB(12 BBs) & 641                              & X/16                                   \\ \hline
TU + DB(10 BBs)  & 616                              & X/8                                    \\ \hline
TU + DB(8 BBs)  & 520                              & X/4                                    \\ \hline
TU + DB(6 BBs)  & 412                              & X/2                                    \\ \hline
TU + DB(4 BBs)  & 315                              & X                                      \\ \hline
output conv     & c                                & X                                      \\ \hline
\end{tabular}
\label{table_arch_spec}
\end{table}

\subsection{Training Procedure}
DVNet is trained using stochastic gradient descent (SGD) on TensorFlow \cite{abadi2016tensorflow}. Combinations of hyperparameters are used to evaluate network performance and memory limitations. A softmax activation function is applied on the output layer to calculate class probabilities for each pixel:
\begin{equation}
p_c(\mathbf{x}) = \frac{\exp(a_c(\mathbf{x}))}{\left(\sum_{k=1}^{C}\exp{a_k(\mathbf{x})}\right)}   
\end{equation}
where $a_c(\mathbf{x})$ is the logit of the channel $c$ of the output at the position of $\mathbf{x}$. $C$ is the number of classes and $p_c(\mathbf{x})$ is the softmax probability for class $c$. Its output is used in the backpropagation for computing loss and optimizing weights.\\

\vspace{2mm}
\noindent
\textbf{Loss function:} pixel-wise cross entropy and dice coefficient are two functions that are used for loss calculation. Pixel-wise cross entropy is the most common cost function for image segmentation:
\begin{equation}
e_c(\mathbf{x}) = \sum_{C}{g_c(\mathbf{x})log\left(p_c(\mathbf{x})\right)}
\end{equation}
where $g_c(\mathbf{x})$ and $p_c(\mathbf{x})$ are the ground truth and prediction values for class $c$ at position $\mathbf{x}$. $C$ refers to the number of classes and $e_c(\mathbf{x})$ is the cross-entropy error at that position. The dice coefficient is used in semantic segmentation tasks, and leverages overlap between the prediction and ground truth: 
\begin{equation}
D = 1 - \left(\frac{2\hspace{1mm}\abs {\mathbf{P} \cap \mathbf{G}}}{\abs{\mathbf{P}} + \abs{\mathbf{G}}}\right)
\label{eqn_dice_loss}\end{equation}
where $\mathbf{P}$ and $\mathbf{G}$ represent prediction and ground truth masks, $\abs{}$ shows the number of elements, and $\cap$ stands for the common elements in the masks.

\vspace{2mm}
\noindent  
\textbf{Evaluation metrics:} This network is trained and evaluated using the pixel-wise accuracy and the intersection over union (IoU). IoU is computed for each class and evaluate the class wise performance:
\begin{equation}
    IoU_c = \frac{\sum_{V}({p_c}\cdot{g_c})}{\sum_{V}\left((p_c + g_c) - (p_c\cdot{g}_c)\right)}
\label{eqn_iou}
\end{equation}
where $c$ is the current class and $V$ is the set of all voxels. $p_c$ and $g_c$ are voxels from the prediction and ground truth masks that are labeled as class $c$. $(.)$, and $(+)$ are intersection and union functions. We evaluate DVNet performance using global accuracy and mean IoU.

\vspace{2mm}
\noindent   
\textbf{Optimizer:} Trainable parameters are initialized using the Xavier initializer \cite{glorot2010understanding}. Adam optimizer \cite{Kingma2014adam} with the initial learning rate of $1\mathrm{e}{-3}$ optimizes weights to minimize loss and improve accuracy. Learning rate is decreased with a decay rate of $0.97$ after each $500$ iteration.

\subsection{Post Processing}
\label{section_post_processing}
DVNet forms the basis for identifying cellular and microvascular surfaces. An explicit model is required for many practical applications. We adapt the most effective cell localization \cite{saadatifard2018robust} and vascular tracing \cite{govyadinov2018robust} algorithms to segment cell nuclei and identify vascular surfaces and connectivity using the masks generated by the proposed DVNet (Figure \ref{fig_model_config}).
There are two available algorithms that work well in cell localization and vessel tracing from the KESM dataset \cite{saadatifard2018robust, govyadinov2018robust}. The performance of these algorithms is further improved by pre-processing this dataset using the proposed CNN network.

\vspace{2mm}
\noindent
\textbf{Cell localization:}\label{ivote3} iVote3 is a robust cell detection technique for 3D large volume datasets \cite{saadatifard2018robust}.
\begin{figure}[!t]
\centering
\includegraphics[width=0.8\linewidth]{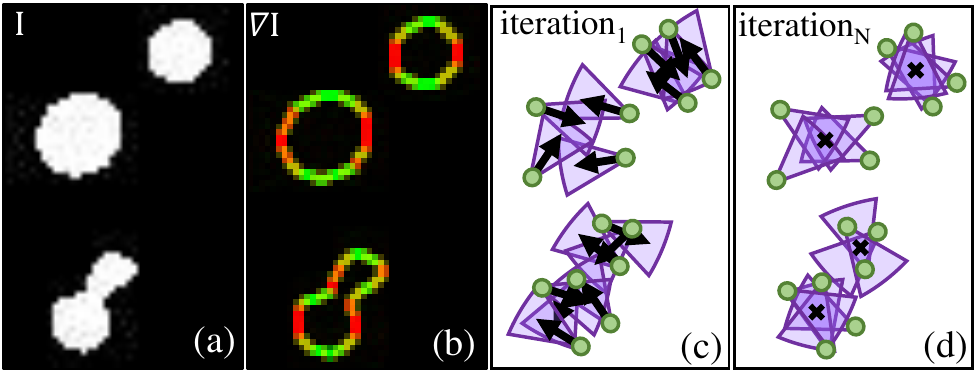}
\caption {Illustrated iVote3 pipeline (a). The input image is created by the neural network.  The gradient of the input image is calculated and used to compute the cell center probabilities for each voxel (b). Arrows show gradient direction, and cones are the vote regions for each voter (shown in green circles) that are defined based on the gradient direction at the first iteration (c). Cones are updated after each iteration, and in the last iteration they converge to the cell center position (d).}
\label{fig_ivote3}
\end{figure}
Iterative voting is an embarrassingly parallel computationally intensive method. To improve performance our implementation is accelerated using GPU shared memory and atomic operations. 
This method is based on the radial symmetry and can detect cells with various sizes. The only input parameter is the maximum radius of the cells in the input dataset. iVote3 computes the gradient of the input, and uses the gradient information to calculate a value for each voxel that represents the probability of being a cell center. Based on the probability mask, the gradient information is refined and a new mask is computed iteratively, until the algorithm converges and outputs a center point for each cell. Figure \ref{fig_ivote3} illustrates an overview of iVote3. This algorithm is fast, requires minimal tuning and shows superior cell detection results on dense, low contrast datasets such as KESM.

\begin{figure*}[ht]\centering
\includegraphics[width=\linewidth]{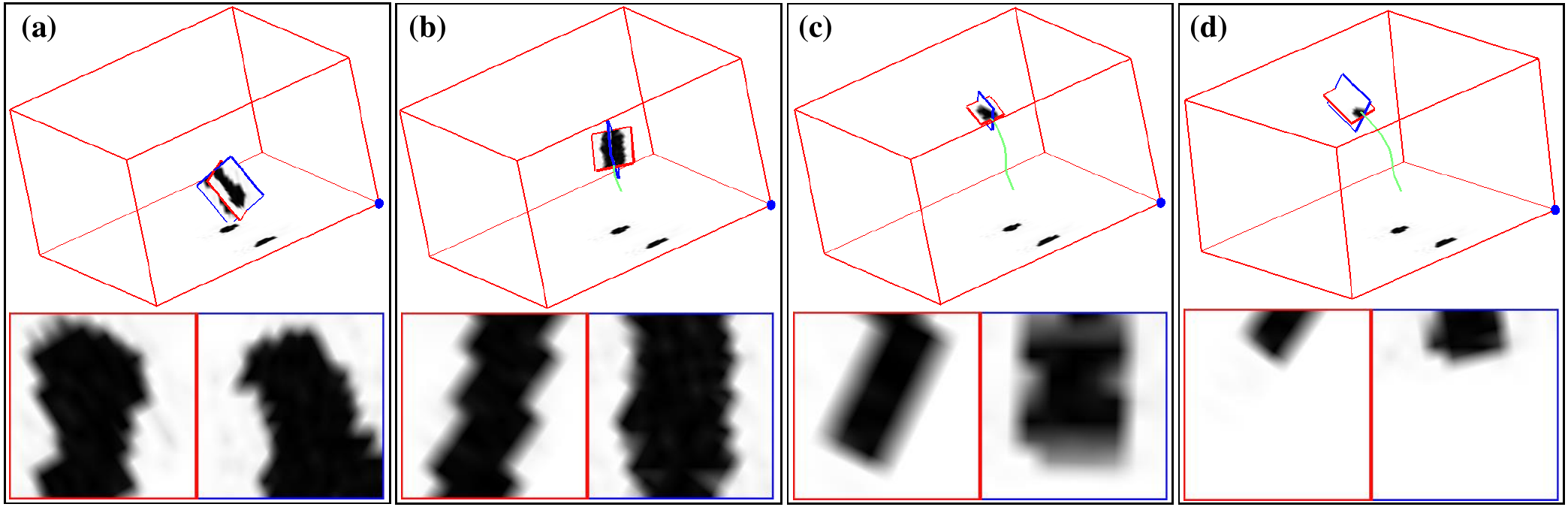}
\caption{Starting with some seed point the algorithm attempts to predict the direction of the fiber (a) by placing two rectangular templates  perpendicular to each other (in red and blue), such that the centerline of the fiber embedded in the tissue is in the exact center (b). The algorithm then continues by advancing along the optimal direction, with an optimal size while tracking previous location along the fiber (c). The algorithm terminates when the fiber disappears (d) or when the algorithm encounters a previously segmented fiber.}
\label{fig:predictor_corrector}
\end{figure*}

\vspace{2mm}
\noindent
\textbf{Centerline Segmentation}\label{vessel_tracking} The segmentation is handled by a  predictor-corrector algorithm that leverages texture sampling to extract centerline, radius and topology information of microvascular network . The algorithm utilizes a set of 2 rectangular templates perpendicular to one another in order to estimate the centerline using multiple steps \cite{govyadinov2018robust}. The process is illustrated in Figure \ref{fig:predictor_corrector}. The final output is a connected graph $G = [\mathbf{V},\mathbf{E}]$, where $v_i$ is the location of a branch point and $e_i$ is a micro-vessel connecting two branch points.

\section{Dataset}\label{sec_KESM_dataset}
\subsection{Data Description}
\begin{figure*}[t]\centering
\includegraphics[width=\linewidth]{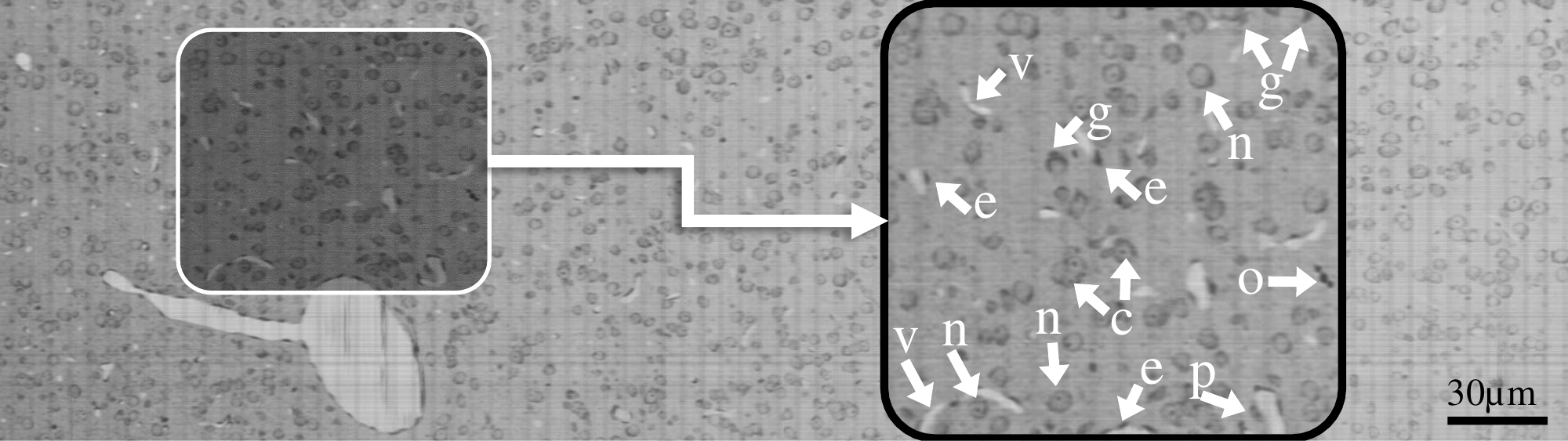}
\caption {Thionine-stained mouse cortex imaged using KESM with a \SI{1}{\micro\meter} section thickness. Thionine is a neucleic acid stain, labeling DNA (cell nuclei) and ribosomal RNA, which is dense within the neuron cytoplasm. The arrows indicate tissue features: endothelial cell nuclei (e), neuron nucleoli (c), glial cells (g), neuron nuclei (n), pericytes (p), and oligodendrocyte nuclei (o). The surrounding neuropil is stained a light grey, making the unstained microvessels (v) visible in 3D.}
\label{fig_kesm_nissl}
\end{figure*}

Knife-Edge Scanning Microscopy (KESM) allows researchers to collect detailed images describing cell structure and vascular/neuronal connectivity across large (\si{\centi\meter^{3}}) volumes \cite{mayerich2008knife}. Optimal stains, such as thionine, provide multiple structural features in a single channel. Thionine staining is common in neuroscience for labeling DNA and ribosomal RNA by binding to acidic proteins and nucleic acids. This label provides contrast for neurons, endothelial cells, and various glial cells. Figure \ref{fig_kesm_nissl} illustrates a cropped section of the thionine-stained mouse cortex imaged using KESM with (\SI{0.6}{\um}\texttimes\SI{0.7}{\um}\texttimes\SI{1}{\um}) resolution.
\subsection{Data Preparation}\label{training_data}
Three disjoint KESM volumes are randomly selected for training and evaluation: 20 $(128\times128\times64)$ voxel volumes for training, and 2 sets of 6 $(64\times64\times32)$ volumes for validation and testing. There is no any overlap area between sets. All volumes are annotated manually to segment cellular and vascular structures. $(64\times64\times32)$ volumes of the training set with their related ground truths form a sample of the input batch. Training batches contain two samples due to high memory requirement of DVNet architecture.\\
\textbf{Data augmentation:} Augmentation is randomly applied to expand training data by randomly cropping and rotating volumes along all three axes.

\section{Experimental Results}
\label{results}
 
Both 2D and 3D implementations of DVNet are trained on the KESM dataset. For 2D training, each image and its ground truth is fed to the network as one input sample. Parameters are selected as: $\theta_\downarrow=0.5$, $\theta_\uparrow=0.3$ and $k=16$ to build the network. Training loss is computed using dice function, and drop out set to $\%10$. The trained modle predictions on the test data is evaluated by IoU per class and accuracy that verifies its performance over other 2D networks (Table \ref{table_compare}).\\
Due to the memory limitation, we build 3D DVNet in three sizes with this combinations: ($\theta_\downarrow=0.3$, $\theta_\uparrow=0.3$ and $k=8$), ($\theta_\downarrow=0.3$, $\theta_\uparrow=0.3$ and $k=16$), and ($\theta_\downarrow=0.5$, $\theta_\uparrow=0.3$ and $k=16$). It is trained with the ($64\times64\times32$) input size and $2$ batch size. Dice and cross entropy functions are examined for computing training loss, and result in the same performance. V-Net \cite{milletari2016v}, and 3D implementation of Tiramisu \cite{jegou2017one} are trained on this dataset to optimize their loss and accuracy on the validation set. To compare the performance of DVNet with state of the art, we used trained models to predict cellular and vascular masks from test set, and evaluate those predictions with manually labeled ground truth (Table \ref{table_compare}).\\
Evaluation results confirmed 3D implementation segment KESM with higher accuracy than 2D implementation. Also among the three sizes, biggest structure performs better while having less trainable parameter that other networks (Table \ref{table_compare}). 

\begin{figure*}
    \centering
    \includegraphics[width=\linewidth]{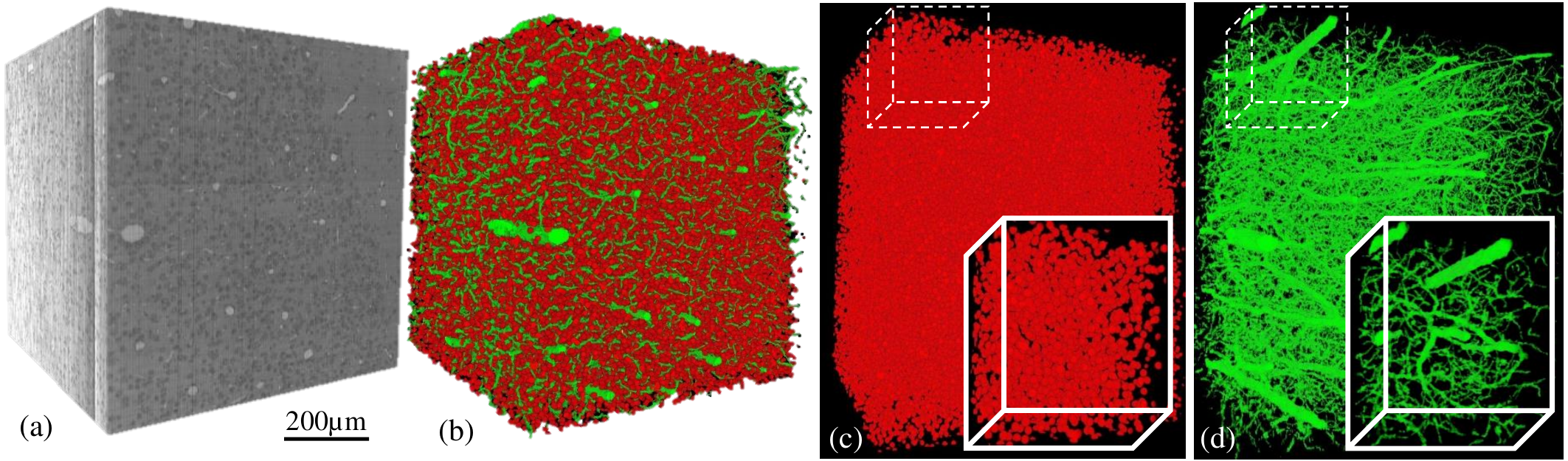}
    \caption{The cellular and vascular structure for $1 mm^3$ region of the nissl stained mouse brain is modeled by the proposed network. (a) The raw volume of the nissl stained mouse brain imaged by the KESM, (b) segmentation results for predicting cellular and vascular structures from the mouse brain, a close view of (c) cellular, and (d) vascular segmentation from the generated model for $1 mm^3$ of the mouse brain.  }
    \label{fig_nissl_cnn}
\end{figure*}

\begin{table*}[b]
\caption{2D and 3D implementation of DVNet ate trained on the KESM dataset. The evaluation of the performance on the test set of DVNet is compared with other 2D and 3D networks.}\label{table_compare}
\resizebox{0.87\columnwidth}{!}{\begin{minipage}{\columnwidth}
\begin{tabular}{llccccccc}
\hline
 & model       & \# parameters (M) & IoU(tissue) & IoU(cell) & IoU(vessel) & mean IoU & accuracy \\ \hline
\hline
\multirow{3}{*}{2D} & UNet\cite{ronneberger2015u} & 31            & 90.5        & 52.8      & 61.1        & 68.2     & 91.5     \\
                    & Tiramisu-103\cite{jegou2017one} & 9.3               & 90.9        & 53.7      & 62.4        & 69        & 91.8     \\
                    & DVNet-v3 \small{$(\theta_D=0.5, \theta_U=0.3, k=16)$}    & 5.3               & \textbf{91.1}        & \textbf{53.8}      & \textbf{63.8}          & \textbf{69.5}      & \textbf{92}       \\ \hline
\multirow{6}{*}{3D} & V-Net \cite{milletari2016v}       & 14                & 92.3        & 62.3      & 61          & 71.9      & 93.1     \\
                    & Tiramisu-67 \cite{jegou2017one}  & 9.5               & 92.5        & 63.2      & 59.6        & 71.8      & 93.3     \\
                    & Tiramisu-103 \cite{jegou2017one} & 26                & 92.7        & 65.2      & 60.9        & 73        & 93.5     \\
                    & DVNet-v1 \small{$(\theta_D=0.5, \theta_U=0.3, k=8)$}    & 2.8               & \textbf{92.9}        & 64.4      & 63.8        & 73.7      & 93.6     \\
                    & DVNet-v2 \small{$(\theta_D=0.3, \theta_U=0.3, k=16)$}    & 9.2               & 92.8        & 64.3      & 64.7        & 74        & 93.7     \\
                    & DVNet-v3 \small{$(\theta_D=0.5, \theta_U=0.3, k=16)$}    & 10.8              & \textbf{92.9}        & \textbf{65.6}      & \textbf{64.8}        & \textbf{74.4}      & \textbf{93.9}     \\ \hline

\end{tabular}
\end{minipage}}
\end{table*}
We used the biggest model of trained DVNet to segment cellular and vascular structure of the KESM dataset. A $1 mm^3$ volume of the KESM mouse brain that includes ($1600\times1500\times1000$) voxels is divided to ($128\times128\times128$) voxels overlapped sections. The trained model predict masks for each subvolumes, and then masks are stitched to model the big region (Figure \ref{fig_nissl_cnn}). 

\begin{figure}
    \centering
    \includegraphics[width=0.5\linewidth]{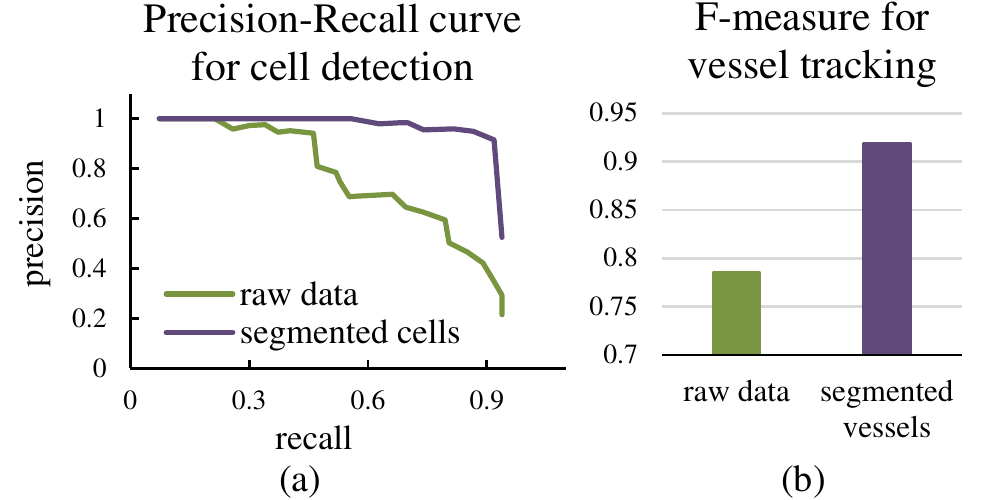}
    \caption{Cellular and vascular models generated by the proposed CNN improve the performance of cell detection , and vascular tracking methods, (a) the mean precision-recall curve computed for cell localization (b) The mean F-score computed for vessel tracking.}
    \label{fig_cell_vessel_pr}
\end{figure}

The 3D model is used to analyze the cellular and vascular architecture in the big region of the brain. To compute the number of cells and their distribution, ivote3 (\ref{ivote3}) is used to localize cell positions. Using the generated model for cell segmentation significantly improves the performance of ivote3 in compare with using the raw images from the KESM (figure \ref{fig_cell_vessel_pr}-a). To analyze the vascular structure in the brain such as computing a capillary distance to another one, average distance between different vessels with different sizes, vessels locations and sizes are needed. The algorithm explained in \ref{vessel_tracking} computes centerlines and radii of vascular network using the vascular model. Figure \ref{fig_cell_vessel_pr}-b indicates the F-measure improvement in comparison with using the raw KESM data. The cell detection and vessel tracking results are compared with a manually labeled ground-truth.

\section{Conclusion}
\label{conclusion}
KESM captures detailed information in individual cell level from the brain architecture, that form a volumetric dataset (\SI{4}{TB}) for a whole (\SI{1}{cm^3}) mouse brain. The diversity in size and shape and low contrast images along with the massive dataset are the challenges for segmenting the brain structure from the KESM images. Memory limitation, time performance and lacking of a robust segmentation algorithm are difficulties that are addressed in this paper.\\
Due to the detail information in the KESM dataset, a deep and densely connected encoder-decoder structure is built to learn semantic features (DVNet). The vanishing gradient problem, which is common in deep networks, is addressed by using the residual function for training network. GPU memory limits the three dimensional models, and it get more confined with dense connections. The memory optimization in DVNet allowed us to train the 3D model. DVNet is trained end-to-end on the KESM dataset to segment cellular/vascular structure of the brain. In compare to other published networks for semantic segmentation, DVNet needs less parameters to be trained and performs better on predicting masks for thionine stained images. Using DVNet to enhance the contrast ofthe thionine stained KESM dataset improves the performance of the cell detection and vessel segmentation algorithms.\\

\section{Discussion and Future Work}
\label{discussion}
DVNet is a deep and densely connected CNN that is computationally intensive and memory limited. The number of trainable parameters and required memory are defined by the number of levels, number of convolutional layer, growth rate, and transition factors. The GPU memory limits the depth of the network to 5 levels in encoder and decoder. It takes \SI{32}{hrs} to train the 3D network on the KESM dataset and achieved 96\% and 93\% training and validation accuracy, and 0.12 and 0.19 for training and validation loss.\\
Hierarchical features extracted at each encoder level and the holistic information computed in the linking unit along with the dense connections in the decoder path of DVNet precisely segment multi-classes from different dataset. It leverages complex spatial features in order to build simplified models of microvascular and cellular structures. DVNet predicts comparable masks on public benchmarks while optimizes memory and the number of trainable parameters.

\section*{Acknowledgment}
This work was funded in part by the National Institutes of Health / National Library of Medicine \#4 R00 LM011390-02 and the Cancer Prevention and Research Institute of Texas (CPRIT) \#RR140013.

\bibliographystyle{unsrt}  
\bibliography{main}

\clearpage
\section*{Supplementary Material}
\subsection*{DVNet results on open source datasets}
\begin{figure}[ht]
\centering
\includegraphics[width=0.5\linewidth]{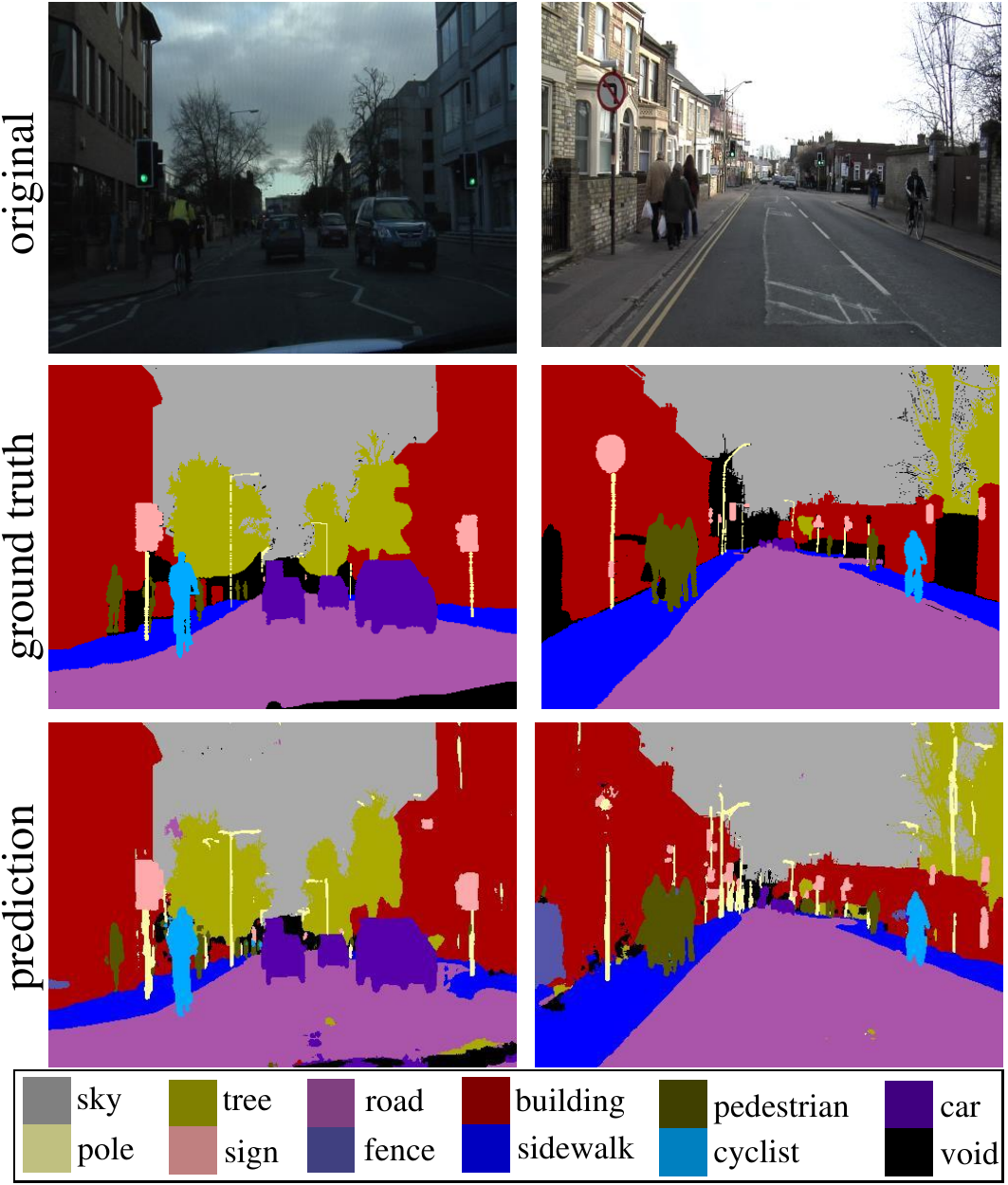}
\caption{CamVid dataset is a 2D urban scene benchmark used for semantic segmentation. Raw images, manually labeled ground truths and segmentation masks created by trained DVNet from the test set are shown.}
\label{fig_camvid_pred}
\end{figure}

We also demonstrate the application of DVNet on open source datasets that have been using in semantic segmentation algorithms. A benchmark of road scenes images (CamVid) have been used to evaluate the performance of 2D implementations. Segmenting the prostate from  background of annotated MRI volumes (Promise2012) is a challenge for 3D structures. \\
\textbf{CamVid dataset (2D)} is a sequence of images from the urban scene videos includes 32 semantic classes \cite{ BrostowSFC:ECCV08, BrostowFC:PRL2008}. We used a set of data from CamVid database created by \cite{badrinarayanan2015segnet} consists of 12 semantic classes. This set includes 367 training, 101 validation, and 233 test RGB images at $(360\times480)$ resolution. \\
2D DVNet is trained on this dataset with $(224\times224\times3)$ input size and $3$ batch size. Horizontal flip and random crop are augmentation methods applied on the training set. Weighted Cross-entropy is used as the loss function to compensate different frequencies of pixels from specific classes in the training set. The network is trained in $50$ thousands iterations using Adam optimizer. The trained model is used to segment test dataset (Figure \ref{fig_camvid_pred}). Intersection over union (IOU) for each class, and overall accuracy computed from DVNet predictions are used to compare the performance with state of the art (Table \ref{table_camvid}).\\

\begin{table}[ht]
\centering
\resizebox{0.9\columnwidth}{!}{\begin{minipage}{\columnwidth}
\caption{Prediction results on the CamVid database generated using 2D DVNet and state of the art. Intersection over union (IOU) per class along with mean IOU and average accuracy is reported to compare their performancec.}
\begin{tabular}{|l|l|l|l|l|}
\hline
model                                                      & Segnet \cite{badrinarayanan2015segnet} & FCN8 \cite{long2015fully} & Tiramisu \cite{jegou2017one} & DVNet \\ \hline
\begin{tabular}[c]{@{}l@{}}parameter \#\\ (M)\end{tabular} & 29.5                                                    & 134.5                                      & 9.4                                           & 5.3   \\ \hline
building                                                   & 68.7                                                    & 77.8                                       & 83                                            & 85.1  \\ \hline
tree                                                       & 52                                                      & 71                                         & 77.3                                          & 91.7  \\ \hline
sky                                                        & 87                                                      & 88.7                                       & 93                                            & 94.8  \\ \hline
car                                                        & 58.5                                                    & 76.1                                       & 77.3                                          & 65.7  \\ \hline
sign                                                       & 13.4                                                    & 32.7                                       & 43.9                                          & 52.7  \\ \hline
road                                                       & 86.2                                                    & 91.2                                       & 94.5                                          & 96.7  \\ \hline
pedestrian                                                 & 25.3                                                    & 41.7                                       & 59.6                                          & 50.4  \\ \hline
fence                                                      & 17.9                                                    & 24.4                                       & 37.1                                          & 67.4  \\ \hline
pole                                                       & 16                                                      & 19.9                                       & 37.8                                          & 29    \\ \hline
sidewalk                                                   & 60.5                                                    & 72.7                                       & 82.2                                          & 88.5  \\ \hline
cyclist                                                    & 24.8                                                    & 31                                         & 50.5                                          & 75.8  \\ \hline
void                                                       & -                                                       & -                                          & -                                             & 24.8  \\ \hline
mean\_IoU                                                  & 46.4                                                    & 57                                         & 66.9                                          & 68.6  \\ \hline
accuracy                                                   & 62.5                                                    & 88                                         & 91.5                                          & 92.4  \\ \hline
\end{tabular}
\end{minipage}}
\label{table_camvid}
\end{table}

\begin{figure}[ht]
    \centering
    \includegraphics[width=0.5\linewidth]{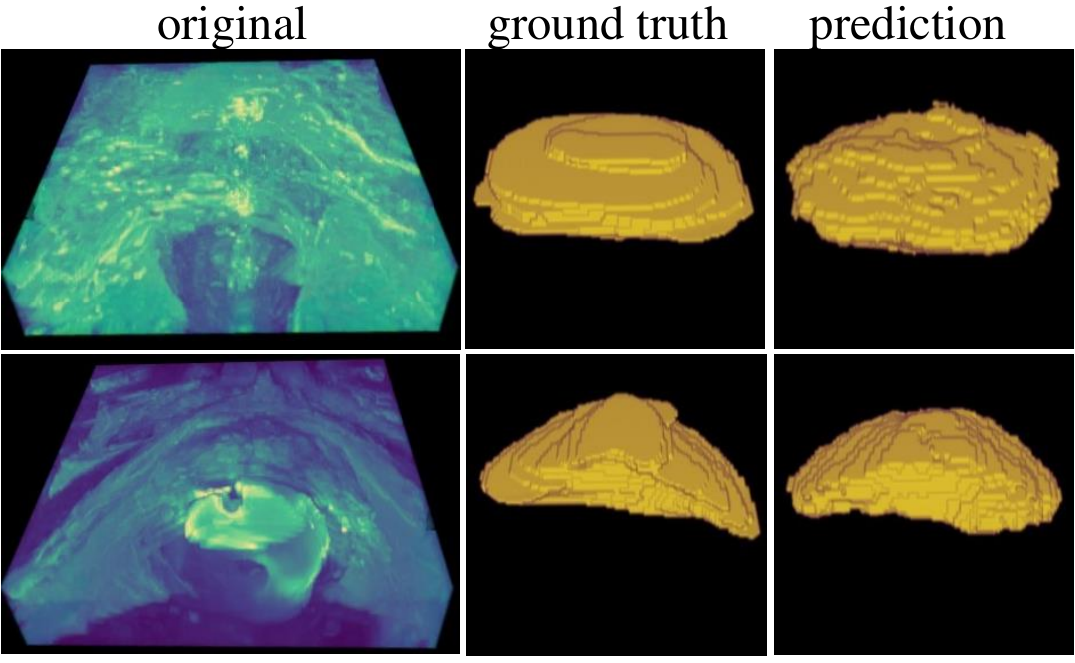}
    \caption{Promise12 is a semantic segmentation challenge on a set of MRI prostate volumes collected from different devices. The prediction masks generated by the proposed network on valid dataset. The network was trained using $\%90$ of the Promise12 dataset, and the $\%10$ of the data, which was not seen by the network during the training process, used for validation.}
    \label{fig_promise_valid}
\end{figure}

\textbf{Promise2012 dataset (3D)} is a challenge on segmenting prostate in MRI volumes collected from different equipment in different hospitals \cite{litjens2014evaluation}. This dataset contains 50 volumes with their manually annotated ground truth used for training and validation sets, and 30 volumes for test set. Since this challenge is still open, labels for the test data are not available. This dataset includes volumes with different sizes and resolutions. $(\%10)$ of the annotated volumes is denoted to the validation set and rest is used for training set. DVNet is trained with $3$ batch size and $(64\times64\times16)$ input size. Data augmentation is applied during the training procedure, and cross entropy is used for loss function. The trained model predicts prostate masks for the validation set in the feedforward manner and with $5$ batch size and $(128\times128\times16)$ input size (\ref{fig_promise_valid}). The average accuracy and mean IOU are computed for prediction on the  validation dataset as $\%96.8$ and $\%88.5$.

\end{document}